\documentclass[aps,floats,floatfix,twocolumn,showpacs,preprintnumbers,prb]{revtex4}

\usepackage{graphicx}
\usepackage{amsmath}
\usepackage{amsfonts}
\usepackage{float}
\usepackage{bm}

\newcommand{\St}{\mbox{\it{St}}} 
\newcommand{\Lo}{\mbox{\it{Lo}}} 
\renewcommand{\Re}{\mbox{\it{Re}}} 

\begin{document}

\title{Clustering of passive impurities in MHD turbulence}

\author{H.\ Homann}
\affiliation{Universit\'e de Nice-Sophia Antipolis, CNRS, Observatoire de
  la C\^ote d'Azur, Laboratoire Cassiop\'ee, Bd.\ de l'Observatoire,
  06300 Nice, France.}
\author{J.\ Bec}
\affiliation{Universit\'e de Nice-Sophia Antipolis, CNRS, Observatoire de
  la C\^ote d'Azur, Laboratoire Cassiop\'ee, Bd.\ de l'Observatoire,
  06300 Nice, France.}
\author{H.\ Fichtner}
\affiliation{Theoretische Physik IV, Ruhr-Universit\"at, 44780 Bochum,
  Germany.}
\author{R.\ Grauer}
\affiliation{Theoretische Physik I, Ruhr-Universit\"at, 44780 Bochum,
  Germany.}

\begin{abstract}
  The transport of heavy, neutral or charged, point-like particles by
  incompressible, resistive magnetohydrodynamic (MHD) turbulence is
  investigated by means of high-resolution numerical simulations. The
  spatial distribution of such impurities is observed to display
  strong deviations from homogeneity, both at dissipative and inertial
  range scales. Neutral particles tend to cluster in the vicinity of
  coherent vortex sheets due to their viscous drag with the flow,
  leading to the simultaneous presence of very concentrated and almost
  empty regions. The signature of clustering is different for charged
  particles. These exhibit in addition to the drag the
  Lorentz-force. The regions of spatial inhomogeneities change due to
  attractive and repulsive vortex sheets. While small charges increase
  clustering, larger charges have a reverse effect.
\end{abstract}


\maketitle

\section{Introduction}
\label{sec:intro}

Many natural and technological settings involve the transport of
particles with mass densities much higher than the main constituents
of the underlying turbulent plasma. For instance in astrophysics,
dense dust grains are suspended in the interstellar
medium\cite{Yan-Lazarian-2003} or in molecular
clouds.\cite{Falceta-Goncalves-etal-2003} Despite usually constituting
only a small fraction of matter in a given astrophysical system, dust
is of central importance for an understanding of many processes, such
as the heating and cooling of interstellar and intergalactic
plasmas.\cite{Tielens-Peeters-2002,Montier-Giard-2004} It is also of
significance for the detection of interstellar magnetic and velocity
fields and their turbulent
properties.\cite{Cruther-etal-2003,Heyer-etal-2008} Dust grains also
play important roles in turbulent protoplanetary disks for the
formation of planets,\cite{Fromang-Papaloizou-2006} for the
interaction of comets with the solar wind,\cite{deJuli-etal-2007} or
the modification of Kolmogorov spectra of weakly turbulent shear
Alfv\'en waves in the ionosphere.\cite{Onishchenko-etal-2003}

In industrial applications, fusion plasmas usually contain a
significant population of heavy charged particles mostly coming from
the introduction of dust in the plasma by erosion of
walls.\cite{tsytovich-vladimirov:1999,Rognlien-2005,Pigarov-etal-2005,
Rosenberg-etal-2008} An understanding of dust dynamics in this context
is not only of interest in order to quantify the effect of impurities
but also to possibly remove the dust particles by exploiting the
notion that scattering through waves can enhance the drag force on
dust particles by orders of magnitude.\cite{deAngelis-2006}

An important aspect of the passive transport of dense particles is the
fact that their inertia can rarely be neglected and is in general
responsible for the apparition of strong inhomogeneities in their
spatial distribution. Indeed, dense particles experience a centrifugal
acceleration in the rotating regions of the flow and tend to
concentrate in those places where strain is dominant. In the purely
hydrodynamical turbulent motion of neutral fluids, this well-studied
phenomenon is usually referred to as \emph{preferential
  concentration}.\cite{squires-eaton:1991} Its main manifestation is
the creation of large particle voids in the most excited parts of the
turbulent flow, which are separated by high-concentration regions
where the particles follow the calmer motions of the fluid.  The
process of ejection from eddies is universal and is also expected in
the case of conducting fluids. It is well known, however, that in MHD
turbulence, the large-strain regions are characterized by the presence
of current/vorticity sheets where both the magnetic field and the
fluid velocity experience strong fluctuations (see, e.g.,
Ref.~\onlinecite{biskamp:2003}).
\begin{figure}[h]
  \begin{center}
    \includegraphics[width=0.5\columnwidth]{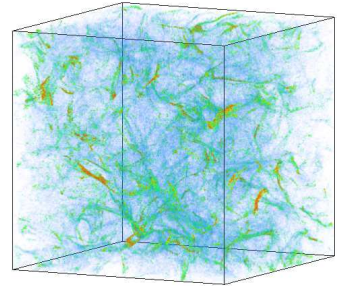} 
  \end{center}
  \vspace{-10pt}
  \caption{\label{fig:mhd_cluster_full}(color online) Density of heavy
  particles in MHD turbulence: gray (blue online): low density, black
  (red online): high density.}
\end{figure}
Consequently particles tend to cluster in the vicinity of these
sheets, as already observed in anisotropic MHD simulations at low
magnetic Reynolds numbers.\cite{Rouson-etal:2008} In statistically
isotropic situations, such an observation is confirmed as illustrated
in Fig.~\ref{fig:mhd_cluster_full} where sheet-like structures can
clearly be observed in the instantaneous three-dimensional
distribution of neutral particles obtained from a direct numerical
simulation.

These correlations between the most active regions of the flow and the
particle distribution contrast much with what is observed in neutral
flows where, conversely, the particles flee coherent structures. This
implies major differences in the properties of dense particles
suspensions depending whether the flow is charged or not. For instance,
we will see that inertial-range particle dynamics in MHD is ruled by
concentration processes rather than ejection processes, leading to
larger fluctuations in their spatial mass distribution.

This paper is organized as follows.  Section~\ref{sec:methods} gives a
detailed description of the model we consider, of its domain of
validity, and of the numerical methods we use to integrate it. In
Section~\ref{sec:prefconc} we present a qualitative comparison of the
effects of preferential concentration in charged and neutral turbulent
flows, emphasizing in each case the different roles played by the flow
coherent structures. Section~\ref{sec:quantitative} is dedicated to a
more quantitative study of particle clustering, both at dissipative
and inertial-range scales. In the former case, we show that particles
cluster on a singular set displaying multi-fractal properties. In
contrast, this scale-invariance does not hold in the inertial range
where the mass distribution rather depends on a scale-dependent
contraction rate of the particle dynamics. In Section~\ref{sec:ions},
we present results on the effect of the particle charge on their
concentration properties. Finally, Section~\ref{sec:conclusion}
encompasses concluding remarks.

\section{Model and methods}
\label{sec:methods}

We consider an electrically conducting fluid whose dynamics is
described by the MHD equations
\begin{eqnarray}
  && \partial_t {\bm u} + \bm u\cdot\nabla\bm u = (\nabla\times{\bm
      B}) \times {\bm B} - \nabla p + \bm f +\nu \nabla^2 {\bm u},\ \
      \ \strut \label{eq:mhd1} \\ && \partial_t {\bm B} =
      \nabla\times({\bm u}\times{\bm B})+ \bm\gamma +\eta_\mathrm{d}
      \nabla^2 {\bm B},\label{eq:mhd2}\\ && \nabla\cdot {\bm u} = 0,
      \quad \nabla \cdot {\bm B} = 0,\label{eq:mhd3}
\end{eqnarray}
where $\bm u$ is the fluid velocity field and ${\bm B}$ is the
magnetic field written with the dimension of a velocity (i.e.\ it is
the magnetic field expressed in the Gaussian system of units divided
by $\sqrt{4\pi\rho_\mathrm{f}}$ where $\rho_\mathrm{f}$ is the fluid
mass density). $\nu$ and $\eta_\mathrm{d}$ are the kinematic viscosity
and the magnetic diffusivity, respectively. $\bm f$ and $\bm\gamma$
are two forces such that the amplitudes of the Fourier modes
associated to wavelengths of modulus less than 2 are kept
constant. Such a choice provides an input of kinetic and magnetic
energies, preventing them to condensate at large length scales. This
means in particular that there is no mean magnetic field, $\langle \bm
B \rangle = 0$, so that the turbulence is statistically isotropic. The
electric field ${\bm E}$ is related to the magnetic field by the Ohm's
law
\begin{equation}
\label{ohmslaw}
{\bm E} = \eta_\mathrm{d} \nabla \times \bm B -{\bm u} \times {\bm B} .
\end{equation}
Here ${\bm E}$ has the dimension of a square velocity, meaning in the
Gaussian system, that it has been multiplied by
$c/\sqrt{4\pi\rho_\mathrm{f}}$.

We consider next a spherical particle embedded in the MHD flow
described above. We assume that its radius $a$ is much smaller than
the smallest active scale of the carrying fluid flow (in turbulence,
the Kolmogorov dissipation scale $\eta = (\nu^3 /
\epsilon_\mathrm{k})^{1/4}$ where $\epsilon_\mathrm{k}$ designates the
mean kinetic energy dissipation rate). We also assume that the mass
density $\rho_\mathrm{p}$ of the particle is much higher than the
fluid mass density $\rho_\mathrm{f}$. These assumptions allows one to
approximate the particle by a point and considering that the main
effect exerted on it by the fluid is a Stokes viscous drag, which is
proportional to the velocity difference between the particle and the
fluid flow.  When in addition the particle is uniformly charged with
density $\rho_\mathrm{c}$ and evolves in an electro-magnetic field, it
is subject to the Lorentz force, so that its trajectory $\bm X(t)$ is
a solution to the Newton equation
\begin{equation}
  \ddot{\bm X} = \frac{1}{\tau}\! \left[\bm u(\bm X\!,t) \!-\!
  \dot{\bm X} \right] + \frac{1}{\ell} \!\left [\bm E (\bm X\!,t)
  \!+\! \dot{\bm X} \!\times \!\bm B (\bm X\!,t) \right]\!,
  \label{eq:particles}
\end{equation}
where the dots stand for time derivatives and $\tau = 2\rho_\mathrm{p}
a^2/(9\rho_\mathrm{f} \nu)$ and $\ell = \rho_\mathrm{p}\,c /
(\rho_\mathrm{c} \sqrt{4\pi \rho_\mathrm{f}})$ (with the charge
density $\rho_\mathrm{c}$ expressed in Gaussian units). Note that we
neglect here the effect of gravity.

The first term on the right-hand side of (\ref{eq:particles})
represents the viscous drag of the particle with the flow. It involves
the \emph{Stokes time} $\tau$, which is the relaxation time of the
particle velocity to the fluid velocity. This characteristic time
scale gives a measure of particle inertia and is usually
non-dimensionalized by the Kolmogorov time scale $\tau_\eta =
(\nu/\epsilon_\mathrm{k})^{1/2}$ to define the \emph{Stokes number}
$\St = \tau/\tau_\eta =
(2/9)(\rho_\mathrm{p}/\rho_\mathrm{f})(a/\eta)^2$. When $\St\ll 1$,
the particle responds very quickly to the flow fluctuations and almost
follows the dynamics of tracers, i.e.\ $\dot{\bm X} \approx \bm u(\bm
X,t)$. Conversely, when $\St\gg1$, a long time is required for the
particle to react and consequently it moves quasi-ballistically with a
low velocity.

The second term on the right-hand side of (\ref{eq:particles})
involves the length-scale $\ell$, which measures the importance of the
Lorentz force exerted by the ambient fluid magnetic and electric
fields onto the particle. By analogy to the Stokes number which
parameterizes the relative motion of particles at scales of the order
of the Kolmogorov dissipative scale $\eta$, a non-dimensional
\emph{Lorentz number} can be introduced as $\Lo =
\tau\,B_\mathrm{rms}/\ell$ (where $B_\mathrm{rms}$ is the root-mean
square (r.m.s.) value of the magnetic field). Note that, conversely to
$\St$, this number involves a large-scale quantity, $B_\mathrm{rms}$,
because of the cross-product quadratic non-linearity appearing in the
Lorentz force.  This number can be written as $\Lo =
(4\sqrt{\pi}/9)\,B_\mathrm{rms}\,\rho_\mathrm{c}a^2/
(c\nu\sqrt{\rho_\mathrm{f}})$ and measures the relative effect of the
Lorentz force with respect to the viscous drag. When $\Lo\ll1$, the
Lorentz force is negligible and inertia dominates. When $\Lo\gg 1$,
the former is dominating the particle dynamics.

%
\squeezetable
\begin{table*}
  \centering
  \begin{ruledtabular}
  \begin{tabular}{ccccccccccccc}
    $\Re_{\lambda}$&$u_\mathrm{rms}$&$B_\mathrm{rms}$&
    $\epsilon_\mathrm{k}$ &$\epsilon_\mathrm{m}$ &
    $\nu=\eta_\mathrm{d}$ & $\delta x$ & $\eta$ &$\tau_\eta$&$L$&$T_L$& $N^3$
    & $N_\mathrm{p}$\\\hline 100 &$0.073$ &$0.14$ &$5.6\cdot
    10^{-4}$&$8.6\cdot 10^{-4}$&$5 \cdot 10^{-4}$&$2.45\cdot
    10^{-2}$&$2.2\cdot 10^{-2}$& 0.94 &2.7& 17 &$256^3$&$10^6$
  \end{tabular}
  \end{ruledtabular}
 \caption{\label{table} Parameters of the numerical simulations.
   $\Re_\lambda = \sqrt{15VL/ \nu}$: Taylor-Reynolds number,
   $u_\mathrm{rms}$: r.m.s.\ velocity, $B_\mathrm{rms}$:
   r.m.s.\ magnetic field, $\epsilon_\mathrm{k}$: mean kinetic energy
   dissipation rate, $\epsilon_\mathrm{m}$: mean magnetic energy
   dissipation rate, $\nu$: kinematic viscosity, $\eta_\mathrm{d}$:
   diffusivity, $\delta x$: grid-spacing, $\eta =
   (\nu^3/\epsilon_\mathrm{k})^{1/4}$: Kolmogorov dissipation length
   scale, $\tau_\eta = (\nu/\epsilon_\mathrm{k})^{1/2}$: Kolmogorov
   time scale, $L = (2/3
   E)^{3/2}/(\epsilon_\mathrm{k}+\epsilon_\mathrm{m})$: integral
   scale, $T_L = L/u_\mathrm{rms}$: large-eddy turnover time, $N^3$:
   number of collocation points, $N_p$: number of particles of each
   species.}
\end{table*}

Note that, besides the hypotheses of small radius and large mass
density that are necessary to write the viscous drag in the present
form, it is implicitly supposed that the particle charge does not
affect the local motion of the charged fluid. For this, it is required
that the Coulomb force exerted by the particle onto the plasma charges
that are at a distance of the order of the particle radius is much
smaller than the Lorentz force exerted on these charges by the ambient
electromagnetic field.  Namely, one must have
$(4/3)\pi\rho_\mathrm{c}a \ll (1/c)\, u_\mathrm{rms} B_\mathrm{rms}\,
\sqrt{4\pi\rho_\mathrm{f}}$. This condition restricts the model
validity to not-too-high values of the particle charge density, and
thus of the Lorentz number $\Lo$. However, some freedom is left for
the choice of parameters. Indeed, the charged particle is
characterized by three parameters: its radius $a$, its mass density
$\rho_\mathrm{p}$ and its charge density $\rho_\mathrm{c}$ and one can
show that it is always possible to find a combination of these
parameters satisfying the three constraints on the validity of this
model with arbitrary values of the Stokes and Lorentz numbers.

Finally, in the proposed model, we neglect the collective effects of
the particles onto the MHD flow, i.e.\ the particles are
\emph{passive}. This implies that the application fields are
restricted to very diluted situations where the particle number
density is so small that their hydrodynamical and electromagnetic
effects on the conducting carrier fluid can be disregarded. This does
not prevent from considering, for instance for statistical purposes, a
large number of such passive particles in the flow.

The numerical simulations are carried out using the reformulation of
the momentum equation (\ref{eq:mhd1}) in terms of the vorticity
$\bm\omega = \nabla\times \bm u$ and a pseudo-spectral solver. The
underlying equations are treated in Fourier space, while convolutions
arising from non-linear terms are computed in real space. A
Fast-Fourier-Transformation (FFT) is used to switch between these two
spaces. The time scheme is a Runge-Kutta scheme of third
order.\cite{shu-osher:1988} The inter-process communication uses the
Message Passing Interface (MPI). The main parameters of the simulation
are given in Tab.~\ref{table}.

In addition to the fluid fields, we introduce three different types of
particles into the flow: fluid tracers corresponding to the degenerate
case $\St = \Lo = 0$, heavy uncharged particles with $\Lo=0$ and 11
different values of $\St$ between $0.05$ and $20$, and finally heavy
charged particles with $\St=1$ and 8 different (positive) values of
$\Lo$ between $0.015$ and $1.5$. One million particles of each species
are uniformly seeded into the statistically stationary flow. After a
period of relaxation of $2.4 \, T_L$, where $T_L$ desigantes the
large-scale eddy turnover time, the statistical analysis of particle
dynamics is started and averages are performed over a time span also
equal to $2.4 \, T_L$.  In order to obtain from the grid values the
velocity and magnetic fields at the particle positions we use a
tri-linear interpolation. This interpolation scheme parallelizes
efficiently and provides sufficient
accuracy~\cite{homann-dreher-etal:2007} for the statistical properties
under consideration.

\section{Preferential concentration in conducting flow}
\label{sec:prefconc}

We focus in this Section on the effect of the Stokes drag on the
qualitative properties of particle clustering. For this we consider
uncharged particles (with $\Lo=0$), so that the Lorentz's force in the
Newton equation (\ref{eq:particles}) can be neglected.  In an
incompressible flow, the dynamics of tracer particles ($\St=0$), whose
trajectories are solutions to $\dot{\bm X} = \bm u(\bm X,t)$, is
volume preserving. Consequently, an initially uniform distribution of
tracers remains uniform at any later time. Because of their drag with
the fluid, heavy particles have a dissipative dynamics and their
spatial distribution tends to develop strong spatial inhomogeneities,
which are observed to span almost all scales of the fluid turbulent
motion (see, e.g., Fig.~\ref{fig:mhd_cluster_full}). The particles
strongly deviate from the fluid motion in those places where the
latter experiences spatial or temporal fluctuations.  The clustering
properties are thus essentially coupled to the statistical properties
of velocity gradients (and accelerations). At small scales (below the
Kolmogorov scale $\eta$ where the flow is differentiable) the spatial
distribution of particles is settled by the one-point fluid gradient
statistics. At inertial-range scales, it is the structure of the
gradient field that determines the spatial organization of particles.
Hence, depending on the observations scales, different processes are
invoked to describe clustering.

The small-scale clustering can be mostly explained using the formalism
developed for dissipative dynamical systems. As is well known
trajectories of such systems tend to concentrate onto a fractal set
called attractor (see, e.g.,
Ref.~\onlinecite{Eckmann-Ruelle:1985}). For heavy particles whose
trajectories solve a non-autonomous second-order differential
equation, this set is embedded in the position-velocity phase space
$(\bm X,\dot{\bm X})$ and evolves dynamically. The instantaneous
positions of particles are obtained by projecting this set onto the
position subspace. The fractal properties of the attractor can be
preserved by this projection, leading to non-trivial scaling laws for
the particle distribution at small scales. The fractal properties of
the attractor are fully determined by the statistical properties of
the linearized particle dynamics,\cite{bec-gawedzki-horvai:2004} which
only involves the fluid velocity gradient along particle paths. The
conducting character of MHD turbulent flow does not alter such a
mechanism.  Although the probability density function (PDF) of
velocity gradients in MHD quantitatively differs from that observed in
neutral fluid turbulence, the qualitative picture of dissipative-scale
clustering is the same.  For instance, in both cases, concentration
processes are optimal for Stokes numbers of the order of unity. Such
qualitative similarities are confirmed in next Section.

\begin{figure}[H]
  \includegraphics[width=0.48\columnwidth]{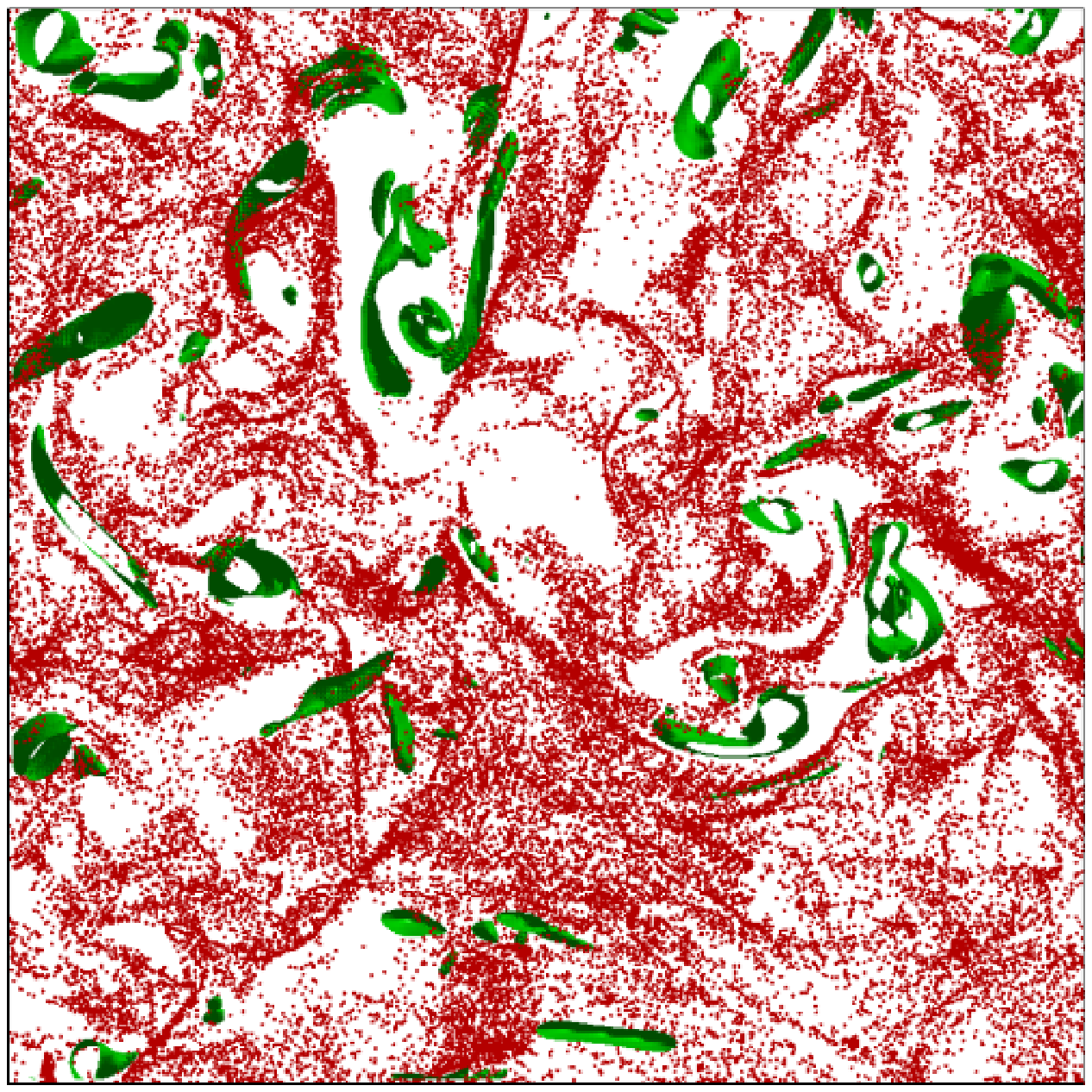} 
  \hfill
  \includegraphics[width=0.5\columnwidth]{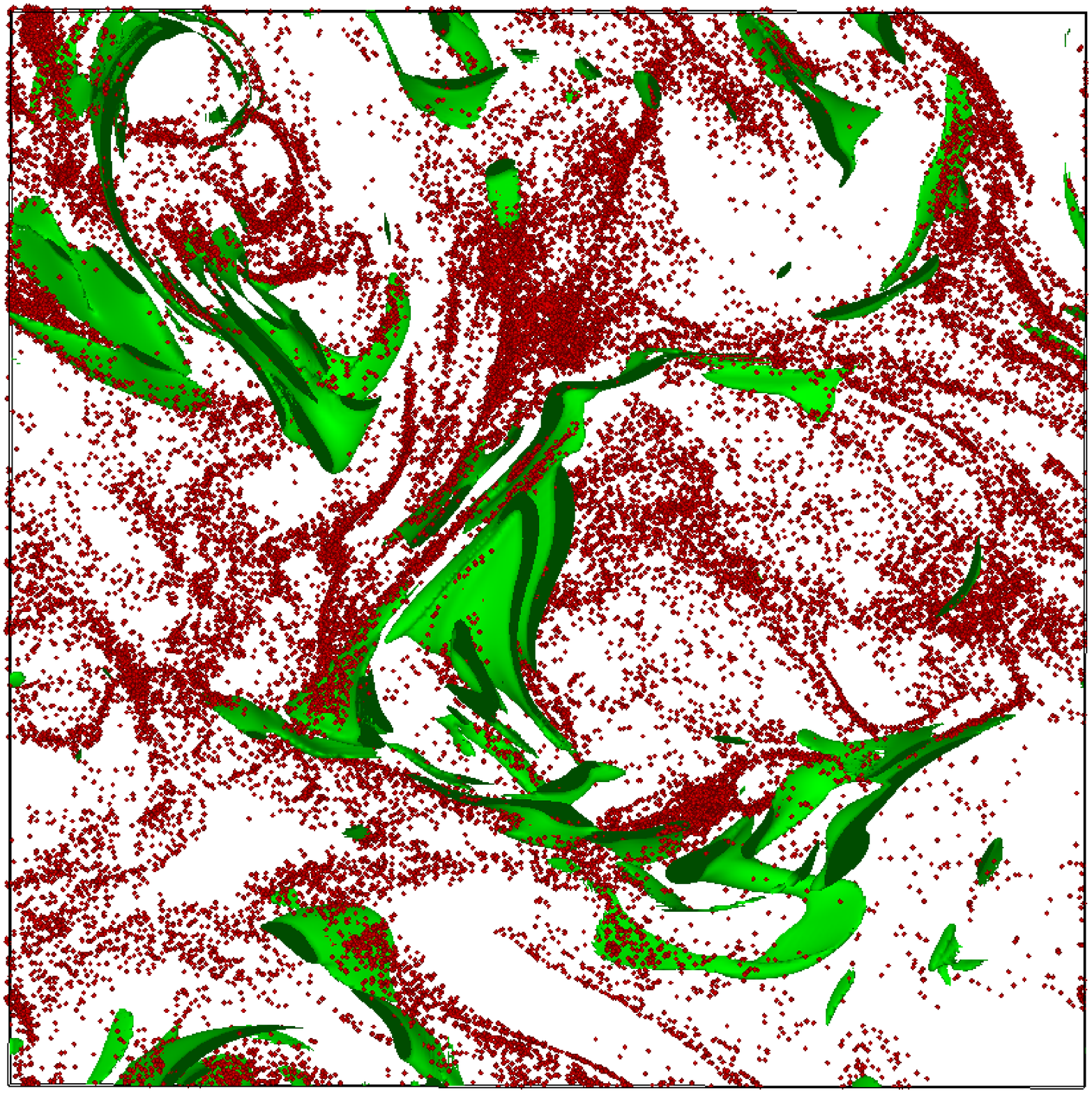}
  \vspace{-8pt}
  \caption{\label{fig:two_clusters} (color online) Isosurfaces of
    vorticity (online: green) and particles (online: red) in a slice
    in purely hydrodynamic turbulence (Left) and in
    magnetohydrodynamic turbulence (Right). In both cases, the domain
    has approximatively the size $256\,\eta\times 256\,\eta$ and the
    slice width is $\approx5\eta$}
\end{figure}
\begin{figure}[b]
  \centerline{
    \includegraphics[width=0.45\columnwidth]{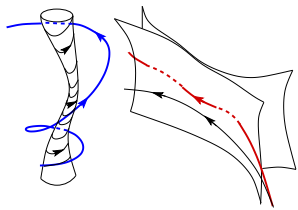}}
  \vspace{-10pt}
  \caption{\label{fig:sketchcluster} Sketch of the motion of a heavy
  particle (bold line) in a vortex filament (left) and in a vorticity
  sheet (right); reference fluid tracer trajectories are drawn as thin
  lines.}
\end{figure}

Strong qualitative differences between neutral and conducting carrier
flows show up in the spatial distribution of particles at scales
within the inertial range.  In hydrodynamical turbulence this
distribution anti-correlates with the spatial organization of the
rotating structures. As illustrated in Fig.~\ref{fig:two_clusters}
(Left) rotating structures where vorticity is strong tend to expel
particles that then concentrate in the calmer regions in between
eddies.  The mechanism of ejection from vortex filaments is sketched
in Fig.~\ref{fig:sketchcluster}.  The flow structure of a conducting
fluid is very different. In MHD turbulence violent gradients
correspond to high strain regions and not to strong rotations. The
coherent structures are thus vorticity/current sheets and the
particles tend to agglomerate on them as illustrated from
Fig.~\ref{fig:two_clusters} Right.
As sketched in Fig.~\ref{fig:sketchcluster} a heavy particle tends to
approach closer to a vortex sheet than fluid tracers, clearly leading
to high concentration in its vicinity.  Moreover, this phenomenon can
be amplified when heavy particles cross the sheet. As vortex sheets
almost correspond to discontinuities of the velocity field, heavy
particles might hence enter a region with a velocity very different
from that of the fluid. This implies a phase during which the particle
decelerates and which is responsible for the concentration of
particles close to the sheets. Of course, the particle clusters and
the sheets do not have a one-to-one correspondence. The fluid velocity
close to the sheet is parallel to it and thus entrains the particles
to the edges of the structure, a picture that applies also to fluid
tracers.\cite{homann-grauer-etal:2007} The fraction of particles that
are close to a given current sheet depends on the dynamical history of
the latter and in particular on its lifetime and on the fact whether
or not it has experienced violent accelerations.

\begin{figure}[H]
    \centerline{
      \includegraphics[width=0.85\columnwidth]{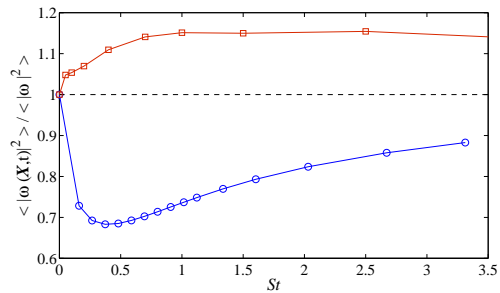}}
    \vspace{-10pt}
    \caption{\label{fig:omega_rms} Vorticty second-order moment along
      particle paths normalized by the Eulerian moment, as a function
      of Stokes number.  {\tiny ${\square}$}: MHD turbulence; $\circ$:
      hydrodynamic turbulence.}
\end{figure}
This different sampling of high-vorticity structures by heavy
particles in hydrodynamic or MHD turbulence has important impact on
the distribution of vorticity along their paths. This can clearly be
observed by measuring vorticity fluctuations at the particle
positions. Figure \ref{fig:omega_rms} shows the second-order moment of
vorticity as a function of the particle Stokes number. While in
hydrodynamic turbulence one observes for Stokes numbers of the order
of unity a large drop-off of more than 30\% with respect to the Eulerian
vorticity, a gain of 15\% can be measured in MHD turbulence. In both
cases this trend displays a maximum for $\St$ of the order of unity
and tends to disappear at large inertia. The effect of inertia on
vorticity sampling is not visible in the core of the vorticity
distribution. It is rather due to a change in the time fraction that
the particles spend in the violent regions with large fluctuations of
the fluid velocity gradient.

\section{Quantitative study of clustering}
\label{sec:quantitative}
We now turn to more precise and systematic ways to characterize
particle clustering. The particle spatial distributions are of
different natures if the fluid velocity field is differentiable or
not. Dissipative and inertial-scale clustering are thus distinguished
here.
\subsection{Small-scale clustering}
At those scales where the fluid velocity field is differentiable,
particle clustering stems from a competition between the stretching
imposed by the fluid velocity gradient and the linear
relaxation/dissipation due to the particle viscous drag. This
materializes by the convergence of particle trajectories to a
dynamically evolving fractal attractor in the position-velocity phase
space. The particle locations are then obtained by projecting such a
singular set onto the configuration space. Hence, depending on the
fractal dimension, the projection is or is not also singular, leading
to possibly fractal spatial distribution of particles.

A (multi)fractal distribution is characterized by a dimension spectrum
$D_p$ (see, e.g., Ref.~\onlinecite{Eckmann-Ruelle:1985}), which is
related to the scaling behavior of the various moments of the
coarse-grained mass distribution. Consider the coarse-grained density
$\rho_r$ defined as the mass of particles inside a fixed box of size
$r$, divided by the volume of the box. The $p$-th order moment of the
coarse-grained density behaves as $\langle \rho_r^p \rangle \propto
r^{(p-1)\,(D_p-d)}$ for $r\ll\eta$, where $d$ denotes the space
dimension. $D_1$ is usually refered to as the \emph{information}
dimension and $D_2$ as the \emph{correlation} dimension. We focus here
on second-order statistics, because they relate to the probability of
finding two particles close to each other and, thus, to estimate the
rate of local binary interactions (as e.g.\ collisions, chemcal
reactions, etc.). Indeed, one can easily show that the probability
$P_2^<(r)$ of finding two particles closer than a given distance $r$
behaves at small scales as $P_2^<(r)\propto r^{D_2}$ (see, e.g.,
Ref.~\onlinecite{Eckmann-Ruelle:1985}).

\begin{figure}[H]
  \centerline{
    \includegraphics[width=0.85\columnwidth]{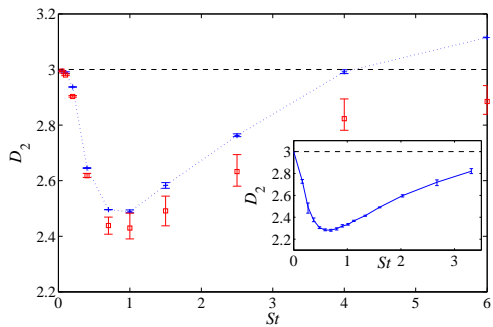}}
  \vspace{-10pt}
  \caption{\label{fig:d2_heavy} Correlation dimension for heavy,
    uncharged particles in MHD turbulence as a function of the Stokes
    number. {\tiny$\square$}: average local slope of $P_2^<(r)$,
    {\tiny $+$}: phase-space correlation dimension $\tilde{D}_2$
    obtained from the fit $P_2^<(r) \simeq Ar^{\tilde{D}_2} +
    Br^{3}$. Inset: same for hydrodynamic turbulence (from
    Ref.~\onlinecite{bec-biferale-etal:2007}).}
\end{figure}
Figure~\ref{fig:d2_heavy} represents the behavior of the correlation
dimension $D_2$ as a function of the Stokes number for neutral
(uncharged) particles. $D_2 $ is observed to decrease from $d=3$ at
$\St=0$, where the dynamics of particles recover that of uniformly
distributed tracers in an incompressible flow. It reaches a minimum
for $\St\simeq 1$ (maximum of small-scale clutering), and increases
again to reach $3$ at large $\St$ where the dynamics of particles
becomes less and less influenced by the flow and approaches the
ballistic motion of free particles. In order to estimate the
correlation dimension, two different methods were used. The first one
consists in averaging the local slope (i.e.\ the logarithmic
derivative) of $P_2^<(r)$ over a scaling range of roughly two decades
in $r$. The error is then estimated through the maximum and minimum
deviations of the local value of the slope from its average. This
method leads to over-estimate clustering for $\St\gtrsim 1$. Indeed,
as was observed in synthetic flows,\cite{bec-cencini-hillerbrand:2007}
the probability of finding in the physical space two particles at a
distance less than $r$ behaves as $P_2^<(r)\simeq A r^{\tilde{D}_2} +
B r^d$, where $\tilde{D}_2$ is the \emph{phase-space} correlation
dimension. When $\tilde{D}_2<d$, the physical-space correlation
dimension is $D_2=\tilde{D}_2$. Conversely, when $\tilde{D}_2>d$, the
particle distribution is space filling (to second order) and
$D_2=d$. Assuming such a form for the two leading terms in $P_2^<(r)$
gives much better estimates of the correlation dimension and allows
one to observe a saturation of the physical-space correlation
dimension to $d=3$ for $\St\ge\St_{\mathrm{cr}} \simeq 4$. As seen
from the inset of Fig.~\ref{fig:d2_heavy}, the correlation dimension
as a function of the Stokes number has a similar behavior as in purely
hydrodynamic turbulence. The main difference is a rescaling of the
horizontal axis by a constant factor and comes from the definition of
the Stokes number, which involves the turnover time associated to the
Kolmogorov dissipative scale. The measurement of the latter as the
typical scale of the velocity gradient gives some uncertainty that
might explain this discrepancy.

The scaling behavior of the coarse-grained density distribution, which
is related to the dimension spectrum, implies a specific
large-deviation form for the PDF of $\rho_r$. Indeed,
one can easily see that if the moments behave as $\langle \rho_r^p
\rangle \propto r^{(p-1)\,(D_p-d)}$, the local dimension $h_r = \log
\rho_r / \log r$ has to follow at small scales $r$ the large deviation
form $p(h_r) \propto \exp[-(\log r)\, H(h_r)]$, where $H(h_r) =
\max_p[(p-1)\,(D_p-d)-p\,h_r]$ (see, e.g.,
Ref.~\onlinecite{bec-gawedzki-horvai:2004}). The rate function $H$ is
positive and convex and attains its minimum (equal to 0) for
$h_r=D_1$.

\subsection{Inertial-range clustering}
\begin{figure}[H]
  \centerline{ \includegraphics[width=0.85\columnwidth]{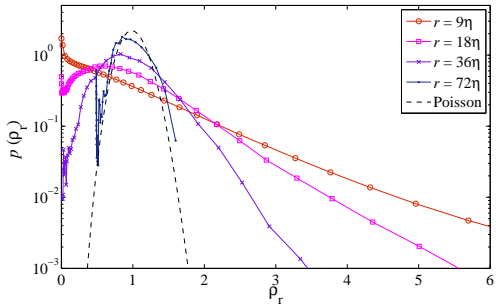}}
  \vspace{-10pt}
  \caption{\label{fig:pdfrho_heavy_varsize} PDFs of the coarse-grained
    particle density over a scale $r$ for neutral particles with
    Stokes numbers $\St=1$ and various values of $r$ as labeled.}
\end{figure}

\begin{figure}[H]
  \centerline{
    \includegraphics[width=0.85\columnwidth]{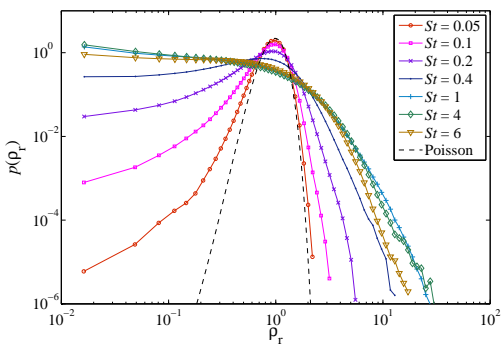}}
  \vspace{-10pt}
  \caption{\label{fig:pdfrho_heavy} PDFs of the coarse-grained
    particle density over a scale $r=18\eta$ for neutral particles
    with several Stokes numbers.}
\end{figure}

The large-deviation form for the distribution of the coarse-grained
density that is discussed above gives a very precise meaning for the
scaling behavior of the fluctuations of mass at small scales. Above
the Kolmogorov scale, where the flow is itself almost self-similar and
spans a large range of timescales, such an invariance is broken
because particles tend to preferentially select those eddies of the
flow whose associated turnover time is of the order of their response
time. Figure \ref{fig:pdfrho_heavy_varsize} represents the various
PDFs of the coarse-grained density $\rho_r$ at fixing the particle
Stokes number and varying the box size $r$ inside the inertial range
of the turbulent carrier flow. One observes that the deviation from a
Poisson distribution is maximal at small scales and that the particle
distribution approaches uniformity when $r$ increases. It is clear
from this picture that the PDF associated to two different scales do
not differ only through a scaling factor, so that there is no large
deviation principle ruling the mass distribution in the inertial
range.
The PDF of the coarse-grained density has noticeable characteristics
that give insights on the mechanisms leading to particle
clustering. Both tails (at low and high values of the mass) are much
wider than what would be expected from a uniform distribution. This
implies that, on the one hand, it is pretty probable to observe a
region with a very large mass, but on the other hand it is also very
probable to observe a region almost empty of particles. The
large-density tails decrease exponentially or slower. This is in
contrast to what was observed in hydrodynamic turbulence, where the
decrease is clearly faster than
exponential.\cite{bec-biferale-etal:2007,bec-chetrite:2007} This
exhibits a difference in the clustering process and might be traced
back to the distinct role of the coherent structures discussed in
previous section. The behavior of the small-density tails is far from
being completely settled. As seen from Fig.~\ref{fig:pdfrho_heavy}, it
seems to display a power-law decay, but the data seems spoiled by a
plateau due to the limited number of particles used in the
simulation. For hydrodynamic turbulence where a larger number of
particles were used, this effect was not so visible and clear evidence
of a power-law behavior could be
obtained.\cite{bec-biferale-etal:2007}

\begin{figure}[H]
  \centerline{
    \includegraphics[width=0.85\columnwidth]{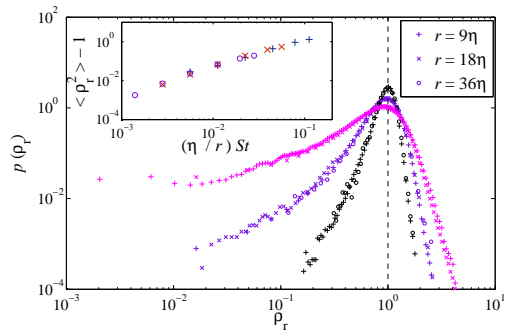}}
  \vspace{-10pt}
  \caption{\label{fig:variance_heavies} Probability distribution of
    the coarse-grained density for three different values of the
    contraction factor $\Gamma = (\eta/r)\,\St$ (widest distribution:
    $\Gamma \simeq 0.07$, intermediate $\Gamma \simeq 0.024$, and
    narrowest: $\Gamma \simeq 0.006$). Inset: deviations from
    uniformity of the variance of the coarse-grained density as a
    function of $\Gamma$.}
\end{figure}

As seen from Fig.~\ref{fig:pdfrho_heavy}, a uniform distribution is
also recovered if, in place of increasing $r$ while keeping $\St$
fixed, one fixes $r$ and decreases $\St$. Strong deviations from
uniformity are indeed observed for $\St\simeq 1$ and the distribution
approaches that of Poisson when $\St\to0$. This leads to asking the
question whether or not the two limits of small inertia
$(r/\eta)\to\infty$ and $\tau\to0$ are equivalent, or more precisely
if the probability distribution $p(\rho_r)$ depends in this limit on a
single parameter $\propto \St/(r/\eta)^\alpha$. To answer this
question, one first notices that in the limit of small inertia, the
particle velocity is to leading order\cite{Maxey:1987}
\begin{equation}
  \dot{\bm X} \simeq \bm v(\bm X, t) = \bm u (\bm X, t) -
  \tau\,(\partial_t\bm u+\bm u \cdot \nabla\bm u) (\bm X, t).
  \label{eq:maxey}
\end{equation}
This approximation consists in assuming that the particles behave as
if they were advected by a synthetic compressible flow. The important
time scale that measures the strength of particle inertia with
response time $\tau$ at a given scale $r$ is given by the inverse of
the rate $\gamma(\tau,r)$ at which a particle blob of size $r$
contracts when advected by this synthetic compressible flow, namely
\begin{equation}
  \gamma = \frac{1}{r^3} \int_{|\bm x|<r} \!\!\!\!  \nabla\cdot {\bm
    v} \,\,\mathrm{d}^3 x = -\frac{\tau}{r^3} \int_{|\bm x|<r}
  \!\!\!\!  \nabla\cdot (\bm u \cdot \nabla \bm u) \,\,\mathrm{d}^3 x.
\end{equation}
The idea is now to understand phenomenologically how the integral on
the right-hand side behaves as a function of $r$. One can first notice
that using the divergence theorem, the volume integral can be
rewritten as the surface integral on $|\bm x| = r$ of $\bm u \cdot
\nabla \bm u$. We next assume that the blob overlaps a vortex/current
sheet where the velocity almost experiences a jump. This implies a
behavior of the surface integral $\sim r^2 u_{\mathrm{rms}} (u_\eta /
\eta)$, where $u_\eta$ is the typical velocity difference over the
Kolmogorov scale $\eta$. In terms of the Stokes number $\St =
\tau/\tau_\eta$, the blob contraction rate thus behave as $\gamma \sim
\St\,u_{\mathrm{rms}} / r $ and is thus proportional to the
non-dimensional contraction rate $\Gamma = (\eta/r) \St$. Finally,
this approaches predicts that the inertial-range distribution of
particles in the limit $\St\to0$ or $r/\eta\to\infty$ depends only on
the non-dimenional contraction rate $\Gamma = (\eta/r) \St\ll1$. As
seen from Fig.~\ref{fig:variance_heavies}, numerical results confirm
this behavior.

\section{Effect of particle charge}
\label{sec:ions}

Despite cases were impurities can be handled as neutral particles in a
conducting fluid,\cite{Fromang-Papaloizou-2006} there are several
situations where the particles have a charge density different from
the carrier flow.\cite{Yan-Lazarian-2003,Falceta-Goncalves-etal-2003}
In this section we investigate the clustering properties of charged
particles arising from a non-vanishing Lorentz number $\Lo$.  To shed
light on the effect of a charge density different from the surrounding
flow, we focus on ensembles of particles with identical Stokes number
but differing Lorentz numbers.

\begin{figure}[H]
  \centerline{
    \includegraphics[width=0.8\columnwidth]{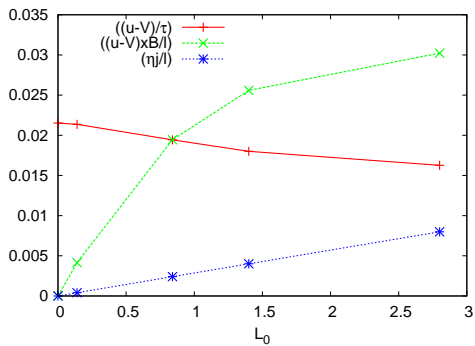}}
  \vspace{-10pt}
  \caption{\label{fig:rms_ions} Rms-contributions of the different terms
    in (\ref{ionForce}) to the force of charged particles.}
\end{figure}

One first notices that Ohm's law (\ref{ohmslaw}) leads to rewrite the
equation (\ref{eq:particles}) for the particle dynamics as
\begin{equation}
  \label{ionForce}
  \ddot{\bm X} = \frac{1}{\tau}[\bm u(\bm X,t) -\dot{\bm X}] +
  \frac{1}{l}([\dot{\bm X}-\bm u(\bm X,t)]\times\bm B + \eta\bm j).
\end{equation}
To get an impression of the relative importance of the three terms
appearing on the right-hand side, Figure \ref{fig:rms_ions} depicts
their root-mean-square values as a function of the Lorentz number. For
small $\Lo$ the drag force dominates. Around $\Lo=0.8$ the
electromagnetic part wins over the drag force.

\subsection{Current sheet selection}
\begin{figure}[H]
  \centerline{
    \includegraphics[width=0.85\columnwidth]{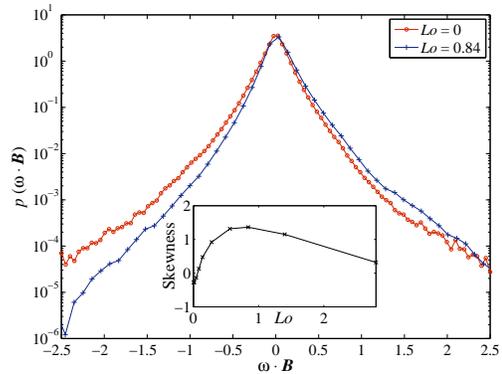}}
  \vspace{-10pt}
  \caption{\label{fig:wb} Probability density of the scalar product
    ${\bm\omega}\!\cdot\!{\bm B}$ sampled along particle paths. Inset:
    skewness $S$ of ${\bm\omega}\!\cdot\!{\bm B}$.}
\end{figure}

We first report some results on the effect of the charge, and thus of
the Lorentz force acting on the particles, on their clustering
properties. While neutral particles tend to cluster in the vicinity of
any current/vortex sheet, charged particles are observed to
concentrate preferentially on a subset among these sheets. Indeed,
because of the Lorentz-force $\frac{1}{l}([\dot{\bm X}-\bm u(\bm
  X,t)]\times\bm B)$ exerted on the particles, there are attractive
and repulsive sheets. In general the velocity of heavy particles is
lower than the fluid velocity at the particle position. Hence,
positively charged particles are attracted by sheets with $\bm
\omega\cdot \bm B > 0$. Figure~\ref{fig:wb} represents the probability
density of this scalar product along the trajectories of neutral and
charged particles with the same Stokes number. One can observe that
the distribution is tilted towards positive values.

A quantitative measure of the preferential selection of certain sheets
is provided by the skewness $\langle (\omega\cdot\bm B)^3
\rangle/\langle (\omega\cdot\bm B)^2\rangle^{3/2}$ of the distribution
of the scalar product $\omega\cdot\bm B$. As is shown in the inset of
Fig.~\ref{fig:wb} the maximal skewness and hence the maximal
preference for attractive sheets is attained for $\Lo\simeq 1$. There
are two possible reasons for the decrease in the preferential
selection of the sheets beyond this value. A first reason relies on a
decreasing Larmor radius ( $\simeq 4\eta$ for $\Lo=0.84$), since the
argument presented for the attraction and repulsion only holds for
Larmor radii much larger than the Kolmogorov-scale. A second reason
originates from a reduction of the degree of concentration at large
Lorentz numbers, as will be discussed in the following two
subsections.

\subsection{Small-scale clustering}

\begin{figure}[H]
  \centerline{
    \includegraphics[width=0.85\columnwidth]{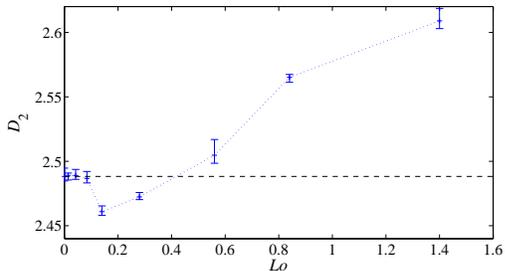}}
  \vspace{-10pt}
  \caption{\label{fig:d2_ions} Correlation dimension for ions}
\end{figure}

As for the neutral particles we measured the correlation dimension
$D_2$ (see Fig.~\ref{fig:d2_ions}). In order to have a visible effect
the Lorentz number has to exceed a certain threshold. Beyond this
relatively flat region the charge leads to an increased concentration
below the Kolmogorov scale, hence to a smaller $D_2$. This is in
relation to the preferential selection of sheets discussed in the
previous subsection. The clustering is enhanced as the attraction
argument provides an additional mechanism for building concentrations.

However, larger Lorentz numbers weaken the preferential concentration
significantly. This can be explained by the fact that the principal
clustering process arises from the drag force. As the electromagnetic
force wins over the drag force for large Lorentz numbers its influence
is weakened and the correlation dimension exceeds that of the neutral
particles, therefore reducing the small-scale clustering.

\subsection{Inertial-range clustering}

\begin{figure}[H]
  \centerline{
    \includegraphics[width=0.8\columnwidth]{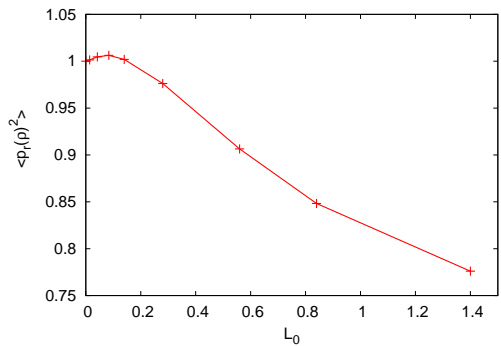}}
  \vspace{-10pt}
  \caption{\label{fig:moment_2_ion} Second order moments of the coarse
    grained density over a scale $r=18\eta$, as a function of the
    Lorentz number $\Lo$, normalized by its value at $\Lo=0$.}
\end{figure}

In the previous subsection we showed that small $Lo$ increase the
clustering at scales below the Kolmogorov scale. Turning to the
implications of a finite Lorentz number for the inertial range of
scales the qualitative findings are
comparable. Fig.~\ref{fig:moment_2_ion} confirms the tendency to an
increased clustering for small charges also in the inertial range,
with again a decrease at higher charges.
Note that the rescaling of the inertial-range distribution as a
function of the contraction factor $\Gamma = (\eta/r)\St$ observed for
neutral particles does not apply in the case of charged particles. The
presence in that case of the Lorentz force, in addition to the viscous
drag, does not allow an approach similar to that presented in
Section~\ref{sec:quantitative}.

\section{Concluding remarks}
\label{sec:conclusion}

We have presented results on the study of heavy particle transport by
MHD turbulent flows. Because of their inertia such particles were
observed to display strong inhomogeneities in their spatial
distribution. More precisely, heavy particles tend to concentrate in
the vicinity of the current sheets that are present in the conducting
flow. Neutral (uncharged) particles approaches sheets, irrespectively
of the nature of the latter, while charged particles, because of a
combined effect of the Stokes drag and the Lorentz force, select those
sheets where the scalar product between the magnetic field and the
vorticity is of the same sign as their charge.

\begin{figure}[H]
  \centerline{
    \includegraphics[width=0.8\columnwidth]{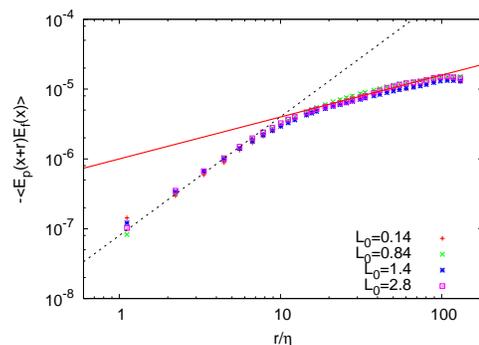}}
  \vspace{-12pt}
  \caption{\label{fig:e_field_correlation} Correlation of the flow
    electric field $E_f$ and the electric field induced by the
    particles $E_p$: dashed (black online) line corresponds to a
    scaling index of 0.6, solid (red online) line to a scaling index
    of 1.7.}
\end{figure}

The particle dynamics considered in this work is entirely passive,
namely they have no feedback on the carrier flow. To estimate their
possible influence on the flow we computed the electric field $\bm
E_p$ produced by the passive charged particles. This field is of
course pronounced in regions with high concentrations. It is oriented
in a way to act against the outer electric field $\bm E_f$ due to the
charge in the flow. To quantify these findings we measured the
correlation $C(\bm r) = \langle \bm E_p(\bm x + \bm r)\bm E_f(\bm x)
\rangle$ and observed that it was negative at all spatial scales. As
seen from Fig.~\ref{fig:e_field_correlation}, this correlation is
observed to behave as a function of the scale like a power-law with
exponent $1.7$ in the dissipative range and $0.6$ in the inertial
range. The origin of these power laws is still an open question which
is clearly related to the concentration properties of the particles.
It seems crucial to include electrical interactions between particles
as soon as the latters are not very diluted in the conducting
flow. Electric repulsive forces should clearly alter both the
small-scale and the inertial-range properties of the particle spatial
distribution but we still expect the preferential selection of current
sheets to occur at inertial-range scales.

Finally we would like to stress a possible application of the results
presented here. Light particles, such as bubble in water, have now
been used for a long time in order to give a direct visualization of
the intense vortex filaments that are present in hydrodynamic
turbulence.\cite{Douady-Couder-Brachet:1991} We have seen here that
heavy particles in magneto-hydrodynamics tend to cluster in the
vicinity of the current sheets that represent the most violent
structures present in the turbulent motion of conducting flows. An
idea could then be to use such particles in order to develop a new
direct visualization technique of MHD flows that is expected to be
very useful for grabbing qualitative and quantitative global
information in turbulent dynamo experiments.

\subsection*{Acknowledgments}
Access to the IBM BlueGene/P computer JUGENE at the FZ J\"ulich was
made available through project HBO22. Part of the computations were
performed on the ``mesocentre de calcul SIGAMM''. The work of H.Homann
benefitted from a grant of the DAAD. This research was supported in
part by the National Science Foundation under Grant No.\ PHY05-51164,
by the Agence Nationale de la Recherche under Project DSPET, DFG-FOR
1048, and the SOLAIRE Network (MTRN-CT-2006-035484).

\end{document}